\documentclass[11pt]{article}
\usepackage{graphics,epsfig,subfigure}
\usepackage{diagbox}

\usepackage{amsthm}
\usepackage{amscd}
\usepackage{amstext}
\newcommand{\ba}{\begin{array}}
\newcommand{\ea}{\end{array}}

\let\w=\omega

\def\be{\begin{equation}}
\def\ee{\end{equation}}
\def\ba{\begin{array}}
\def\ea{\end{array}}

\def\dalemb#1#2{{\vbox{\hrule height .#2pt
        \hbox{\vrule width.#2pt height#1pt \kern#1pt
                \vrule width.#2pt}
        \hrule height.#2pt}}}

\newcommand{\bea}{\begin{eqnarray}}
\newcommand{\eea}{\end{eqnarray}}

\def\ocal{{\mathcal{O}}}

\def\charge{{q}}

\begin{document}

\begin{center}
\vspace{1cm} { \LARGE {\bf Excited states of holographic superconductors with backreaction}}

\vspace{1.1cm}

Yong-Qiang Wang$^\flat$ \footnote{yqwang@lzu.edu.cn, corresponding author}, Hong-Bo Li$^\flat$ \footnote{lihb17@lzu.edu.cn},  Yu-Xiao Liu$^\flat$ \footnote{liuyx@lzu.edu.cn},
 and Yin Zhong$^\sharp$ \footnote{zhongy@lzu.edu.cn}

\vspace{0.7cm}

{\it $^\flat$Institute of Theoretical Physics $\&$ Research Center of Gravitation, Lanzhou University, Lanzhou 730000, China }

\vspace{0.5cm}

{\it $^\sharp$School of Physical Science and Technology $\&$ Key Laboratory for Magnetism and Magnetic Materials of the MoE, Lanzhou University, Lanzhou 730000,  China}

\vspace{1.5cm}

\end{center}

\begin{abstract}
\noindent
In this paper we investigate the model of the
anti-de Sitter gravity coupled to a  Maxwell field
 and a free, complex scalar field,
and  construct a fully back-reacted holographic model of   superconductor
with excited states. With the  fixed charge $q$, there exist
a series of excited states of holographic superconductor with the corresponding critical chemical potentials.
The condensates
as  functions of the
temperature for the two operators $\ocal_1$ and $\ocal_2$   of excited states are also studied.
For the optical conductivity in the excited states, we find that there exist  the  additional peaks in the  imaginary and real parts of the conductivity. Moreover,   the number of peaks  corresponding to
 $n$-th excited state is equal to $n$.
\end{abstract}

\vspace{6cm}

\pagebreak

\section{Introduction}
In the past ten years,  the anti-de Sitter/conformal field theory (AdS/CFT) correspondence \cite{Maldacena:1997re,Witten:1998qj,Aharony:1999ti}  has  been used to study the
  strongly correlated systems  in condensed matter physics, and received a great deal of attention.
In the remarkable  papers \cite{Gubser:2008px,Hartnoll:2008vx,Hartnoll:2008kx}, with the
$U(1)$ symmetry breaking in a four-dimensional Schwarzschild-AdS
black hole background, the condensate of a scalar field could be
interpreted as the holographic realization of superconductor condensate. When replacing the scalar field with other matter fields,  one could  obtain the holographic models for various kinds of  superconductors.
For example, by introducing  a $U(1)$ gauge field coupled to a symmetric, traceless second-rank tensor field in the bulk spacetime, the
holographic realization of d-wave supeconductor was obtained in \cite{Chen:2010mk,Benini:2010pr,Kim:2013oba}.
With the breaking  of  $SU(2)$ gauge symmetry, the holographic model of p-wave superconductor was also  studied
in \cite{Gubser:2008wv}. In addition,
two alternative holographic models of p-wave superconductor  could been realized from the condensates of a two-form field \cite{Aprile:2010ge} and  a complex,
massive vector field with $U(1)$ charge \cite{Cai:2013pda,Cai:2013aca}, respectively. Considering  the weak-link barrier
between two superconductor
materials \cite{Josephson:1962zz}, the holographic model of  Josephson junction was studied in \cite{Horowitz:2011dz,Wang:2011rva,Siani:2011uj,Kiritsis:2011zq,Wang:2011ri,Wang:2012yj,Cai:2013sua,Takeuchi:2013kra,Li:2014xia,Liu:2015zca,Hu:2015dnl,Wang:2016jov,Kiczek:2019lmz}.
A top-down construction of holographic superconductor from superstring theory was discussed in \cite{Gubser:2009qm},  and
a similar construction using an M-theory truncation was proposed in \cite{Gauntlett:2009dn,Gauntlett:2009bh}.
For reviews of holographic superconductors,
see \cite{Hartnoll:2009sz,Herzog:2009xv,Horowitz:2010gk,Cai:2015cya}.

 Recently,  the authors in \cite{Wang:2019caf}  constructed  the novel solutions
of holographic superconductor,  in which excited states of a scalar field were explored in the probe limit.
In contrast to the ground state of holographic superconductors,  the excited states could have some nodes  along the radial  coordinator, in which the value of the scalar field could change its
sign. Moreover, there exist
a series of excited states of holographic superconductor with the corresponding critical chemical potentials.
It is  interesting to
mention that the
conductivity in the excited states of holographic superconductors  has an additional pole in imaginary part and a delta function in
real part arising at the low temperature inside the gap. Similar behaviours for  conductivities were found  in the  holographic model of
s-wave superconductor with mass $m^2 = -9/4$ \cite{Horowitz:2008bn} and the
type II Goldstone bosons \cite{Amado:2013xya}. These behaviors indicate  the emergence of   new bound states, which  embodies
 the existence of  excited states.

Away from the large charge limit, the backreaction of the matter fields
on the bulk metric needs to be considered \cite{Hartnoll:2008kx}.
In the present paper,
we study a holographic s-wave superconductor model with full back-reaction, focusing on  the excited states of scalar field. After numerically solving the coupled equations, condensate of scalar field and the optical conductivity will be studied.


The paper is organized as follows. In Sec. \ref{sec2}, we review the gravity dual model of a holographic superconductor with backreaction. We
 explore the
relations between the critical chemical potential and the charge, and study the characteristics of the condensate and  conductivity of the excited states with backreaction  in Sec. \ref{sec3}. The conclusions and discussions are given in the last section.

\section{Review of holographic superconductors}\label{sec2}
In this section, we review the model of a Maxwell field and a charged complex scalar field coupled to Einstein gravity with a negative
cosmological constant $\Lambda$ in the four-dimensional spacetime \cite{Hartnoll:2008kx}. The bulk action reads
\be
\mathcal{S}= \frac{1}{16\pi G}\int \mathrm{d}^4x \sqrt{-g} \left[R+\frac{6}{\ell^{2}}-\frac{1}{4}F^{\mu\nu}F_{\mu\nu}-(\mathcal{D}_\mu \psi)(\mathcal{D}^\mu \psi)^*-m^{2}\psi \psi^*\right],\label{Lagdensity}
\ee
where $F_{\mu\nu}=\partial_{\mu}A_{\nu}-\partial_{\nu}A_{\mu}$ is the field strength of the $U(1)$ gauge field, and $\mathcal{D}_\mu=\nabla_\mu-iq A_\mu$  is
the gauge covariant derivative with respect to $A_{\mu}$.
The constant $\ell$  is the AdS length scale, $m$ and $q$ are the mass and  charge of the complex scalar field $\psi$, respectively.
Due to the existence of the Maxwell and complex scalar fields, the strength of the backreaction of the matter fields on the spacetime metric  could be tuned by the charge $q$.
 The backreaction on the gravity will decrease when $q$ increases, and the large $q$ limit ($q\rightarrow \infty$) corresponds to the probe limit (non-backreaction) of the matter fields. In
this paper  beyond the probe limit, we will numerically solve the full set of equations with
finite charge $q$.
The  equations of the scalar and Maxwell fields can be derived from  the action (\ref{Lagdensity}):
\begin{eqnarray}
(\nabla_{\mu}-i q  A_{\mu})(\nabla^{\mu}-i q A^{\mu})\psi-m^{2}\psi&=&0,\label{scalarequ}\\
\nabla_{\mu}F^{\mu\nu}-i q [\psi^{\ast}(\nabla^{\nu}-i q A^{\nu})\psi-\psi(\nabla^{\nu}+ i q A^{\nu})\psi^{\ast}]&=&0, \label{maxwellequ}
\end{eqnarray}
and variation of the action with respect to the metric leads to the Einstein equations
\be\label{einsteineq}
R_{\mu\nu} - \frac{g_{\mu\nu} R}{2} - \frac{3 g_{\mu\nu}}{\ell^2} = T_{\mu\nu},
\ee
with the stress-energy tensor of the scalar  and Maxwell fields given by
\be
T_{\mu\nu}=\frac{1}{2} F_{\mu\lambda} F_\nu{}^\lambda - \frac{g_{\mu\nu}}{8} F^{\lambda\delta} F_{\lambda\delta} + \frac{g_{\mu\nu}}{\ell^2} \psi\psi^*
\displaystyle{  - \frac{g_{\mu\nu}}{2}(\mathcal{D}_\lambda \psi)(\mathcal{D}^\lambda \psi)^* + (\mathcal{D}_{(\mu} \psi)(\mathcal{D}_{\nu)} \psi)^*  }.
 \ee

 For simplicity, we shall work with units in which $\ell=1$.
 When the charged scalar field $\psi=0$, the solution of Einstein equations (\ref{einsteineq}) is the well-known Reissner-Nordstr\"{o}m-AdS (RN-AdS) black hole. This solution with a planar symmetric horizon can be written as follows
\be \label{metric}
  ds^{2} = -f(r)dt^{2}+\frac{dr^{2}}{f(r)}+r^{2}(dx^{2}+dy^{2}),
\ee
and the gauge field is given by
\be
A=A_t dt= \mu_0 (1-\frac{1}{r})dt,
\ee
where $f(r)=r^2-\frac{1}{r}\left(1+\frac{\mu_0^2}{4}\right)+\frac{\mu_0^2}{4 r^2}$, and $\mu_{0}$ is the chemical potential for $U(1)$ charge. Thus, the event horizon of the black hole is located  at $r=1$ and the boundary of the asymptotical AdS spacetime is at $r\rightarrow0$. The Hawking temperature is given by
\be
T_{RN}=\frac{(12-\mu_{0}^{2})}{16\pi},
\ee
which can be regarded as the temperature of the holographic superconductors. Thus, the solution of a RN-AdS black hole could exist only for $\mu_0  \leq \mu_{ext}=\sqrt{12}$, which
describes the extremal RN-AdS black hole with zero temperature.

It is well known that there exists a critical temperature $T_{c}$, below which the black hole has a scalar hair that breaks the $U(1)$ gauge symmetry spontaneously. As for $T>T_{c}$, the black hole with a scalar hair degrades into  RN-AdS black hole.
In order to build a holographic model of  holographic superconductors, one could introduce
the following ansatz of metric:
\be
\label{eq:metric}
ds^2 = - g(r) e^{-\chi(r)} dt^2 + \frac{dr^2}{g(r)} + r^2 \left(dx^2 + dy^2
\right)
\,,
\ee
and matter fields  with
\begin{equation}\label{ansatz}
  A=\phi(r) dt, \;\;\;\;\; \psi=\psi_{n}(r),\;\;\;n=0,1,2,\cdots\;.
\end{equation}
Subscript $n$ is   the principal
quantum number of the scalar field, and $n = 0$ is regarded as the ground state and $n \geq 1$ as
the excited states.

With the above ansatzs (\ref{eq:metric}) and (\ref{ansatz}),  the independent equations of motion can be
deduced as follows
\bea\label{eq:Efinal2}
\psi_n'' + \left(\frac{g'}{g} -{\chi'\over 2} + \frac{2}{r} \right) \psi_n' + \frac{\charge^2 \phi^2 e^\chi}{g^2} \psi_n -
\frac{m^2 \psi_n}{g}  = 0 \,,\\
\phi'' + \left({\chi'\over 2} + \frac{2}{r} \right)
\phi' - \frac{2 \charge^2 \psi_n^2}{g} \phi = 0 \,,\label{eq:Efinal3}\\
\chi' + r\psi_n'^2  + {r \charge^2 \phi^2\psi_n^2e^\chi\over g^2 }=0 \label{eq:Eone}\,,\label{eq:Efinal4} \\
\frac{1}{2} \psi_n'^2 +  \frac{\phi'^2e^\chi}{4g} + \frac{g'}{g
r}+ \frac{1}{r^2} - \frac{3}{g} +
\frac{m^2\psi_n^2}{2 g} + \frac{\charge^2 \psi_n^2 \phi^2 e^\chi}{2 g^2} = 0 \,.\label{eq:Efinal5}
\eea
To obtain the numerical solutions for the four functions  $\psi_n$, $\phi$, $g$ and $\chi$,
one must set suitable boundary conditions imposed at the event horizon $r=1$:
\be
g=0,\;\;\;\;\; \phi=0.
\ee
In addition, at the asymptotic
boundary $r\rightarrow \infty$, the asymptotic behaviors of the four functions take the following forms
\begin{eqnarray}\label{asympphi}
\psi_n &=& \frac{\psi_n^{(1)}}{r^{\Delta_-}} + \frac{\psi_n^{(2)}}{r^{\Delta_+}} + \cdots \,,\,\,\,\,\,\,
\phi = \mu - \frac{\rho}{r} + \cdots \,,\nonumber\\
g  &=& r^2 +\cdots\,,\,\,\,\,\,\,
\chi =  0+\cdots\,,
\end{eqnarray}
with
\begin{equation}\label{delta1}
\Delta_\pm=\frac{3\pm\sqrt{9+4 m^2}}{2},
\end{equation}
where $\psi^{(i)} (i=1,2) $ are the corresponding expectation values of the dual scalar operators $\mathcal{O}_{i}$.  According to AdS/CFT dictionary, the parameters $\rho$  and $\mu$ are the  charge density and chemical potential in the dual conformal field theory, respectively.
 In  the four-dimensional spacetime,  the values of mass $m$ need to satisfy the Breitenlohner-Freedman (BF) bound ($m^{2}\geq-9/4$) \cite{Breitenlohner:1982bm}.
To simplify, we will set $m^2=-2$ in this paper.
\section{Numerical results}\label{sec3}
In order to numerically solve the above coupled equations (\ref{eq:Efinal2})-(\ref{eq:Efinal5}),
we introduce a
new radial variable $z=1/r$
which maps the semi-infinite region $[1,\infty)$ to the finite region $[1,0]$.  Thus the event horizon and the asymptotic boundary could be fixed at $z = 1$ and $z= 0$, respectively.
We focus on pseudospectral methods based on the Chebyshev expansion, and the number of grid points ranges between
 $60$ to $300$ in the integration region $0 \leq z \leq 1$. The iterative process has been performed by using
the Newton-Raphson method, and the relative error for the numerical solutions in our paper is
estimated to be below $10^{-5}$.

\begin{figure}[h]
\begin{center}
\epsfig{file=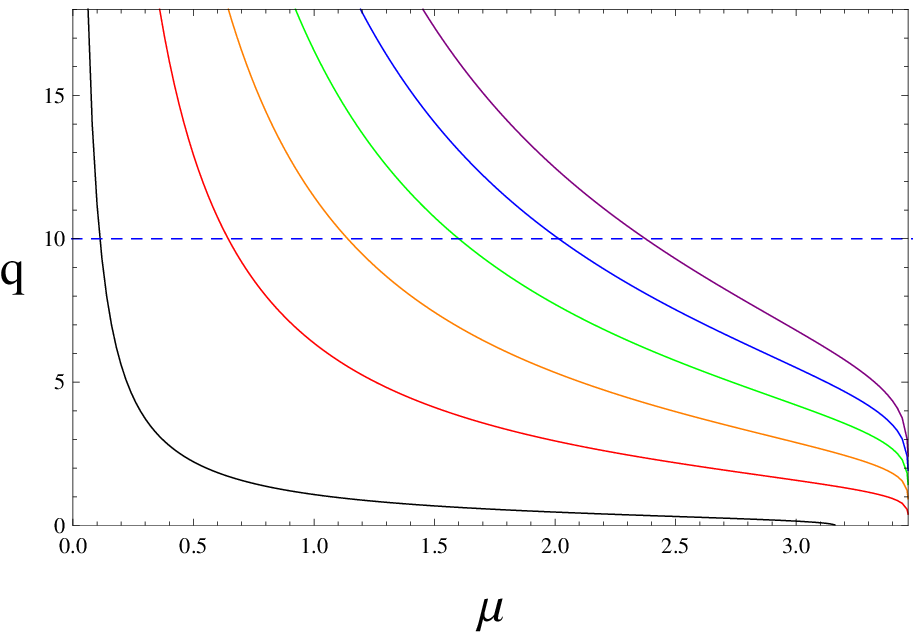,width=2.9in,angle=0,trim=0 0 0 0}
\epsfig{file=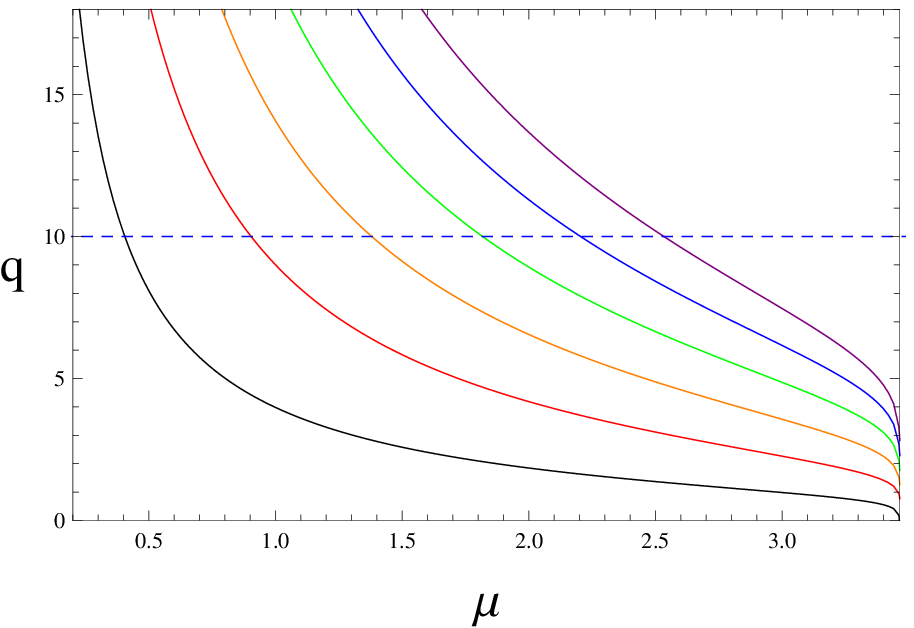,width=2.9in,angle=0,trim=0 0 0 0}
\end{center}
\caption{The charge $q$  as a function of the critical chemical potential $\mu$ for the operators $\ocal_1$ (left) and $\ocal_2$ (right).
The curves from bottom  to top correspond to the cases from the ground to fifth excited states. A typical blue dashed horizonal line with $q = 10$  is  shown to how to obtain the critical
chemical potential of different excited states.
  \label{fig:critical}}
\end{figure}

With the  fixed charge $q$, there exists a critical value of the chemical potential $\mu$,
  below which
the RN-AdS black hole solution is stable and the scalar field vanishes. Above this value, the  charge static solution becomes unstable, the scalar
condensate could form and lead to  the  spontaneously breaking of   $U(1)$ gauge symmetry.
 Before solving the full  dynamic
equations of motion including Einstein equations  numerically, we would adapt the same method as in \cite{Horowitz:2013jaa}, and find the critical value of $\mu$ with the fixed charge $q$. It is convenient to solve the time-independent scalar field equation at a RN-AdS black hole background with different values of chemical potential:
\begin{equation}
(\nabla_\mu \nabla^\mu -m^2)\psi = q^2\,A_\mu A^\mu\,\psi\,,
\label{eq:eige}
\end{equation}
with the mass $m^2=-2$.
The above equation could be recognized as an eigenvalue problem with the eigenfunction $\psi$ and eigenvalue $q^2$.
In a RN-AdS black hole  with the fixed chemical potential background, we could obtain a series of eigenvalues of charge $q$.
 The smallest eigenvalue, that is, zero-mode of eigenvalue, corresponds to  the ground  state, while larger modes of eigenvalues correspond to excited states.

In  Fig.~\ref{fig:critical}, we show the charge $q$  as a function of the critical chemical potential $\mu$ for the operators $\ocal_1$ and $\ocal_2$ in the left and right panels, respectively.  There are six curves on both plots denoting six different kinds of
solutions from the ground to fifth excited states.
In right panel, a typical blue dashed horizonal line with $q = 10$ indicates the  critical chemical potential  $\mu_{c}=$ 0.41 (ground state), $0.90$ (1st-excited), $1.38$ (2nd-excited) and $  1.82$ (3rd-excited), $ 2.20 $  (4th-excited), $2.53$  (5th-excited) for the operator  $\ocal_2$, respectively, which corresponds to the values of $q=10$  shown in  Table \ref{table1}.
For each curve, the charge $q$ decreases with the increasing
chemical potential $\mu$.
We see that for the fixed amplitude $\mu$,  the  charge $q$ of
excited state is larger than the ground state.
	\begin{table}[!htbp]
		\centering
		\begin{tabular}{|c|c|c|c|c|c|c|c|}
			\hline
       \diagbox{$q$}{$\mu_{c}$ }{n} &$0$&$1$&$2$&$3$&$4$&$5$\\
			\hline
			$6$ & 0.67& 1.46&2.15 & 2.67& 3.04& 3.26 \\
              \hline
			$10$ &0.41  & 0.90 &1.38  &1.82  & 2.20  &  2.53  \\
          \hline
          $14$ &0.29  &0.65  & 1.00 & 1.34   & 1.66 & 1.96  \\
          \hline
		\end{tabular}
\caption{The critical chemical potential $\mu_{c}$ for the operator $\ocal_2$ from the ground to fifth excited states with the  charges $q=6, 10, 14$, respectively.}\label{table1}
	\end{table}

In  Table \ref{table1}, we present the
results of the critical chemical potential $\mu_{c}$ for the operator $\ocal_2$ from the ground to fifth excited states with the  charges $q=6, 10, 14$, respectively. It is obvious that an excited state   has a higher critical chemical potential than the ground state, which means the excited state has a higher critical charge density $\rho_c$. Due that the critical temperature is proportional to $\rho^{-1/2}_c$ in four-dimensional spacetime, the excited state has a lower  critical temperature than the ground state , which is similar to the behavior in the probe limit \cite{Wang:2019caf}.

\subsection{Condensates}
\begin{figure}[h]
\begin{center}
\epsfig{file=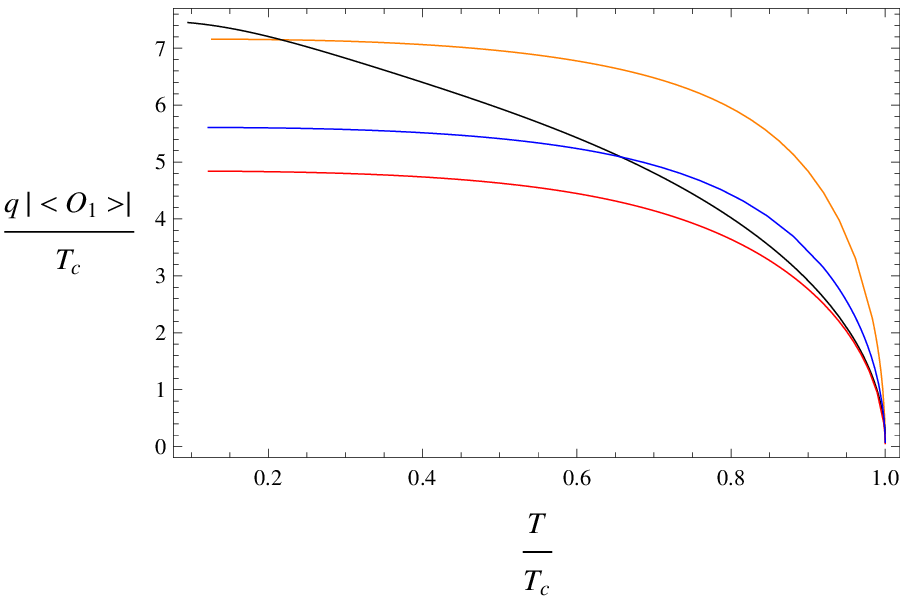,width=2.86in,angle=0,trim=0 0 0 0}
\epsfig{file=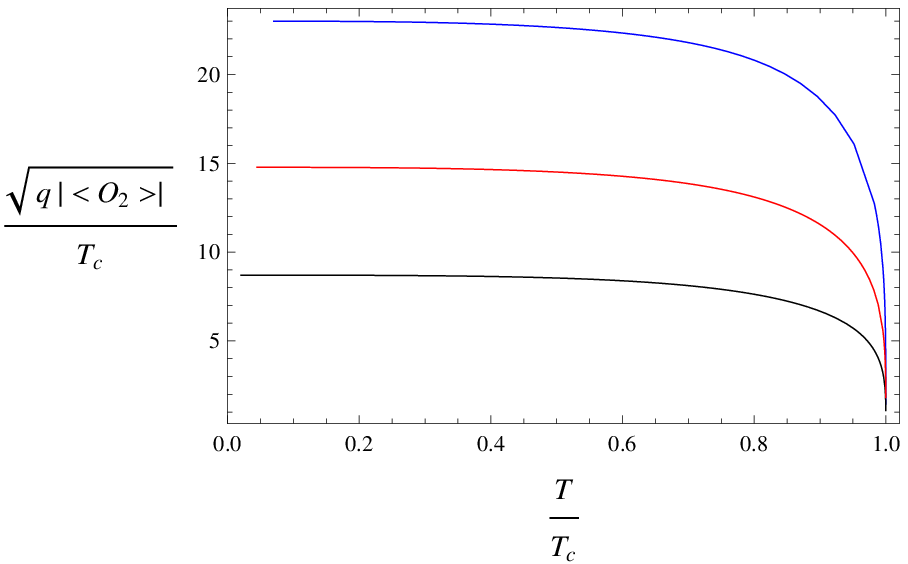,width=3.05in,angle=0,trim=0 0 0 0}
\end{center}
\caption{The  values of condensates as   functions of the
temperature for the two operators $\ocal_1$ (left) and $\ocal_2$ (right) of excited states with the fixed charge $q=6$. In both plots, the black,  red  and  blue lines denote the ground state, first and second states, respectively. In the left panel,  the orange  line denotes   third excited  state.
  \label{fig:condensate}}
\end{figure}

 According to AdS/CFT duality,
 the expectation value of the  condensate operator $\mathcal{O}_{i}$ could be
connected to the scalar field $\psi^{(i)}$:
\be
\langle \ocal_i\rangle = \sqrt{2} \psi^{(i)} \,,\quad i=1,2.
\ee
The values of condensates as  functions of the
temperature for the two operators $\ocal_1$  and $\ocal_2$   of excited states with the charge $q=6$ are shown  in the left  and right panels  of Fig. \ref{fig:condensate}, respectively.
In both plots, the black,  red  and  blue lines correspond to the ground state, first and second states, respectively.
 In the left panel,
the condensate of the ground state for the operator $\mathcal{O}_{1}$  starts to curve upwards as ones approach low
temperatures and appears similar to that of the probe limit. However, the condensate of excited states  appears
to converge as $T\rightarrow0$, and there exists the constant condensate near zero temperature.  Moreover, in contrast to the behavior of condensate for the operator $\mathcal{O}_{1}$  in probe limit \cite{Wang:2019caf}, we find the condensate of the first-excited state is also smaller than that of the ground state in the left panel,
however, it is interesting that the second-excited and higher excited states  have  larger condensate than the first excited state,  which is different from the probe limit case. To clarify this issue, we plot an extra orange curve corresponding to the third excited state.
In the right panel, the condensate of the excited
states for the operator $\mathcal{O}_{2}$  appears to converge as  the temperature $T\rightarrow0$, and the condensate of each excited state is larger than the ground state, which  appears similar to that of the probe limit.

\subsection{Conductivity}
In this subsection we study the optical conductivity of the excited states of holographic superconductors with backreaction.  According to the holographic duality,
we can calculate electromagnetic fluctuations  in the bulk  geometry.
 Considering  a time
dependence  perturbation of $\delta A= A_x(r) e^{- i \w t} dx$,  the linearized equation of the
Maxwell equation  is given as
\be\label{eq:Axeq}
A_x'' + \left[\frac{g'}{g} - \frac{\chi'}{2} \right] A_x'
+ \left[\left(\frac{\w^2}{g^2} - \frac{\phi'^2}{g} \right) e^{\chi} - \frac{2 \charge^2
\psi_n^2}{g} \right] A_x  =  0 \,. \label{eq:ax}
\ee
When implementing ingoing wave boundary conditions at the horizon,
we could obtain the asymptotic behaviour of the gauge field $A_x$ at the boundary:
\be
A_x = A_x^{(0)} + \frac{A_x^{(1)}}{r} + \cdots.
\ee
With the AdS/CFT correspondence and  Ohm's law,  the conductivity can be computed by the following formula
\be\label{eq:conductivity}
\sigma(\w) = - \frac{i A_x^{(1)}}{\w A_x^{(0)}} \,.
\ee
In Eq. (\ref{eq:Axeq}),  the   scalar field  $\psi_0$ corresponds to the ground state, and  the conductivity of the ground state   has been studied in \cite{Hartnoll:2008kx}.

\begin{figure}[h]
\begin{center}
\epsfig{file=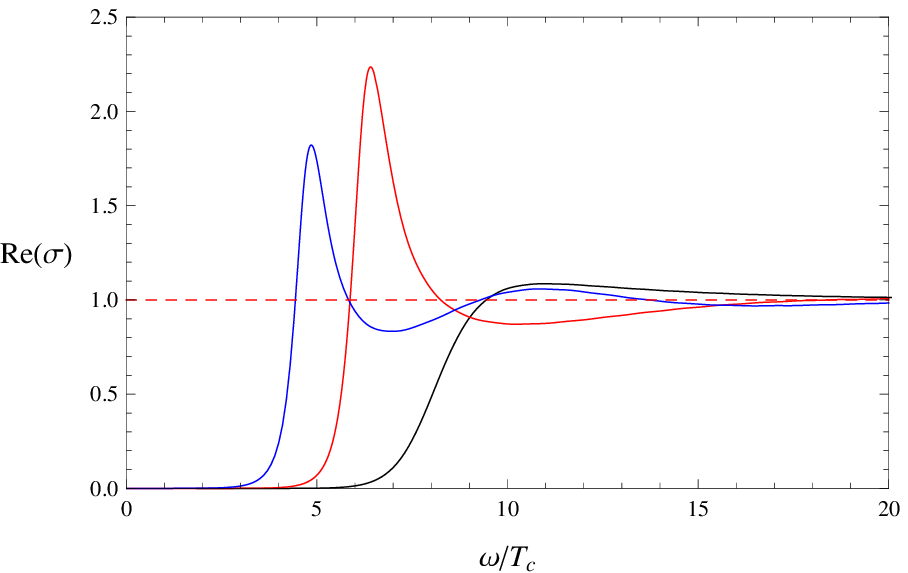,width=2.94in,angle=0,trim=0 0 0 0}
\epsfig{file=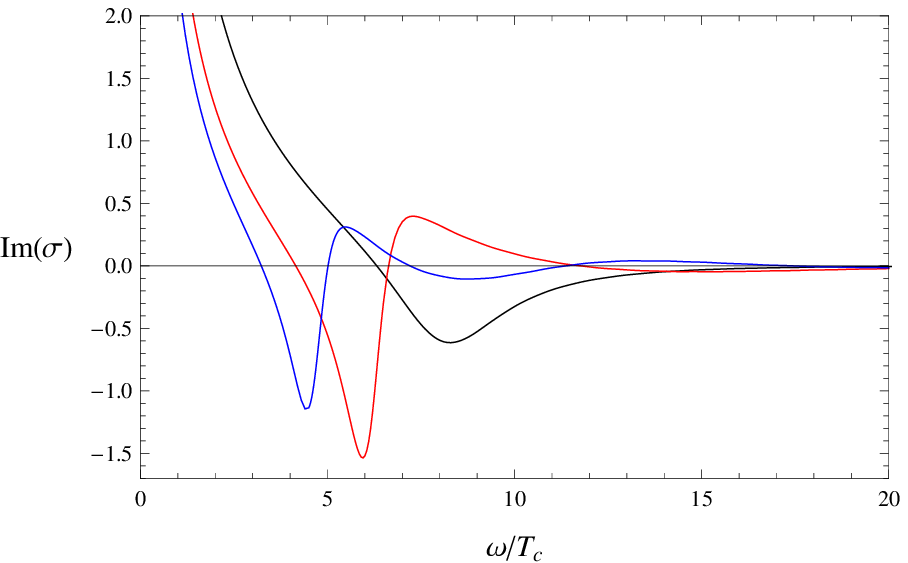,width=3.01in,angle=0,trim=0 0 0 0}
\end{center}
\caption{ The real and imaginary parts of optical conductivity as   functions of the frequency at the temperature $T/T_c\simeq0.2$.
In both plots, the black,  red  and  blue lines denote the ground state, first and second states, respectively.
  \label{fig:gap}}
\end{figure}
In Fig. \ref{fig:gap},  we present the AC conductivity as a function of the
frequency $\omega$  at the low temperature $T/T_c\simeq0.2$ for the  operator $\ocal_2$  from the ground state to  second excited state.
The real and imaginary parts of conductivity are plotted in the left  and right panels, respectively.
 In both plots, the black,  red  and  blue lines correspond to the ground state, first and second states, respectively.
From the imaginary part  in the right panel,  we could see the optical conductivity of excited states also  develops a gap at certain frequency
  which can be identified  as the minimum of
the imaginary part of the optical conductivity \cite{Horowitz:2008bn}. Moreover,
the gaps of excited states are located  inside the gap of the ground state.
It is interesting to note that there exists an additional peak in the  imaginary part of the conductivity
for the first excited state, and  two additional peaks in  $\text{Im}[\sigma]$ of the
 second excited state. Similar behaviours  also appear  in  the real part of the conductivity.
 In the left panel of Fig. \ref{fig:gap},  there also  exist   additional peaks in the  excited states. Moreover,   the number of peaks
 corresponding to
 $n$-th excited state is equal to $n$.
  Recent work  in \cite{Wang:2019caf} shows that
in the probe limite case,  there exists an additional pole in
$\text{Im}[\sigma]$ and a delta function in $\text{Re}[\sigma]$  arising at low temperature inside the gap.
 In contrast to  the probe limit case, we find that due to the introduction of backreaction, the pole and delta function can be  broaden into  the peaks with finite width.

\section{Discussion}
In this paper, we numerically solved the full  dynamic
equations of motion including Einstein equations,  and   constructed the holographic model of  excited states of
holographic superconductors with backreaction. When the temperature drops below the critical temperature,  the condensate of holographic superconductor begins to appear, which could  be regarded as the ground state solution.
Further decreasing the temperature  to another  critical temperature, a new branch of solution with a node in radial direction begins to develop, which is the first excited state solution. As the temperature continues to decrease to lower values, we  could obtain a series of excited states of holographic superconductor with corresponding critical temperatures, which is similar to the probe limit case.
We also studied the optical conductivity in the excited states of holographic superconductors with backreaction.
In contrast to the conductivity of the excited states in the probe limit case,
 it is very interesting to note  that due to the introduction of backreaction, the pole and delta function in optical conductivity at  the probe limit case  can be  transformed into  the peaks with finite values at the backreaction case.  Moreover,   the number of peaks
 corresponding to
 $n$-th excited state is equal to $n$.

There are some interesting extensions of our work which we plan to investigate in future
projects. First,
in order to break the translational
symmetry  which means  the momentum of  charge carriers could be  dissipated, the authors in \cite{Horowitz:2013jaa} introduced   a gravitational background lattice to the holographic model.
It would be
very interesting to study the optical conductivity of
the excited states of holographic superconductor with holographic lattices.
The second extension of our study is
to explore the properties of
holographic superconductors with backreaction using  the semi-analytical techniques \cite{Siopsis:2010uq}.
Finally, we are
planning to investigate the excited states of the p-wave  holographic superconductor with backreaction
and study the excited vector   condensates and optical conductivity  in
future work.

\subsection*{Acknowledgements}
We would like to thank Jie Yang and Li Zhao  for  helpful discussion.  Parts of
 computations were performed on the   shared memory system at  institute of computational physics and complex systems in Lanzhou university. This work was supported by the National Natural Science Foundation of China (Grants
No. 11875151, No. 11522541 and No. 11704166).

\end{document}